\documentclass[%
reprint,
 showpacs,
amssymb,
 aps,
pra,
 longbibliography
]{revtex4-1}

\usepackage{amsmath,dsfont}
\usepackage{mathbbol}
\usepackage{bbold}
\usepackage{mathrsfs}
\usepackage{graphicx}
\usepackage{dcolumn}
\usepackage{bm}
\usepackage[usenames]{color}
\usepackage[normalem]{ulem}

\usepackage[per-mode=symbol]{siunitx}   
\usepackage[version=3]{mhchem}
 
\DeclareSIUnit\intensity{\watt\per\centi\meter\squared}
\DeclareSIUnit\fieldstrength{\volt\per\centi\meter}
\DeclareSIUnit\kfieldstrength{k\volt\per\centi\meter}
\DeclareSIUnit\energy{cm^{-1}}

\newcommand{\melement}[3]{\ensuremath{\left\langle #1 \left|#2\right|#3\right\rangle}}%
\newcommand{\degree}{\ensuremath{^\circ}}%
\newcommand{\ie}{i.\,e.}%

\newcommand{\singletp}{\ensuremath{{}^1\text{P}}}

\newcommand{\doubletseven}{\ensuremath{{}^2\text{S}^e}}
\newcommand{\doubletpodd}{\ensuremath{{}^2\text{P}^\text{o}}}

\newcommand{\pzero}{\ensuremath{\mathcal{P}_0}}
\newcommand{\pone}{\ensuremath{\mathcal{P}_1}}
\newcommand{\ptwo}{\ensuremath{\mathcal{P}_2}}

\newcommand{\ket}[1]{\left|#1\right\rangle}

\newcommand{\expected}[1]{\left\langle #1\right\rangle}
\newcommand{\mtot}{\ensuremath{M_\text{tot}}}

\newcommand{\uam}{\affiliation{Departamento de Qu\'imica, M\'odulo 13, Facultad de Ciencias, Universidad Aut\'onoma de Madrid, 28049 Madrid, Spain}}%
\newcommand{\aarphys}{\affiliation{Department of Physics and Astronomy, Aarhus University, 8000  Aarhus C, Denmark}}%

\begin{document}

\title{Photoionization of aligned excited states in neon by attosecond laser pulses}

\author{Juan J.\ Omiste}\email{juan.omiste@uam.es}\uam
\author{Lars Bojer Madsen}\aarphys

\date{\today}
\begin{abstract} 
We describe numerically the ionization process induced by linearly and circularly polarized XUV attosecond laser pulses on an aligned atomic target, specifically, the excited state Ne$^*(1s^22s^22p^5[\doubletpodd_{1/2}]3s[^1\text{P}^o])$. We compute the excited atomic state by applying the time-dependent restricted-active-space self-consistent field (TD-RASSCF) method to fully account for the electronic correlation. We find that correlation-assisted ionization channels can dominate over channels accessible without correlation. We also observe that the rotation of the photoelectron momentum distribution by circularly polarized laser pulses compared to the case of linear polarization can be explained in terms of differences in accessible ionization channels. This study shows that it is essential to include electron correlation effects to obtain an accurate description of the photoelectron emission dynamics from aligned excited states.
\end{abstract}

\maketitle
\section{Introduction}
\label{sec:introduction}

In ultrafast science, non-linearly polarized attosecond XUV pulses or combinations of them, such as bicircular co- and counter-rotating pulses, have been used to produce asymmetries in the photoelectron spectra of isotropic atomic states~\cite{NgokoDjiokap2013a}, including dichroism~\cite{NgokoDjiokap2014} and to induce vortex patterns in photoelectron momentum distributions~\cite{NgokoDjiokap2015}. Aspects of the latter predictions have been confirmed experimentally in the femtosecond multiphoton ioization regime~\cite{Pengel2017}. Very recently, an attosecond precision interferometric method has been developed to measure angle-resolved phases in ultrafast photoemission~\cite{You2020}. Circularly or near-circularly polarized laser pulses, have also played an important role in the near-infrared regime, where they have been used to access a possible time delay in tunneling ionization~\cite{Eckle2008a,Eckle2008b,Pfeiffer2012,Armstrong2020}. Nowadays, non-linearly polarized laser sources are also used to probe the internal structure of non-isotropic quantum systems, such as open shell or excited atoms and molecules.  It is of particular interest to use laser pulses, to distinguish between different enantiomers of a molecule,~\ie, between molecules which are related by a reflection~\cite{Yachmenev2019} and to probe and control chirality with attosecond sources~\cite{Cireasa2015,Ayuso2019,Beaulieu2018}.

Many theoretical studies have been performed to describe atoms and molecules in circularly polarized laser pulses of short duration, in particular in strong fields~\cite{Martiny2009,Barth2014a,Eckart2018a,Kheifets2020} and in the multiphoton absorption domain~\cite{Han2020a,Ivanov2013}. These works introduce approximations to describe the many-electron dynamics. Hence, the single-active-electron (SAE) approximation is often introduced. As an example, the time delay in the photoionization of the excited state of Li$^*(1s^22p[^3\text{P}^o])$ interacting with circularly polarized light has been described in this framework~\cite{Ivanov2013}. However, the SAE breaks down when electron correlation becomes significant, making mandatory to use a more general framework to account for the electron dynamics. Thus, to fully account for the electron correlation we require a methodology capable of capturing the many electron character of the wave function~\cite{Hochstuhl2014}. Among others, let us highlight methods related to configuration interaction (CI)~\cite{Bauch2014} and multiconfiguration time-dependent Hartree-Fock (MCTDHF)~\cite{Meyer1989,Beck2000a,Koch2006,Lin2020,Lode2020}. CI and MCTDHF methods expand the wave function in configurations, where the CI considers the orbitals as time-independent while MCTDHF assumes that not only the amplitude of each configuration is time-dependent, but also the orbitals. This feature gives the MCTDHF method a flexibility that allows the use of a much smaller number of configurations than CI. In particular, we mention a set of methods which aim to take into account only the most relevant configurations, such as time-dependent complete-active-space self-consistent field (TD-CASSCF)~\cite{Sato2013}, time-dependent configuration-interaction singles (TD-CIS)~\cite{Rohringer2006,Greenman2010,Teramura2019} and time-dependent restricted-active-space self-consistent field (TD-RASSCF)~\cite{Miyagi2013,Miyagi2014,Miyagi2014b}. 

\begin{figure}
\centering
    \includegraphics[width=.95\linewidth]{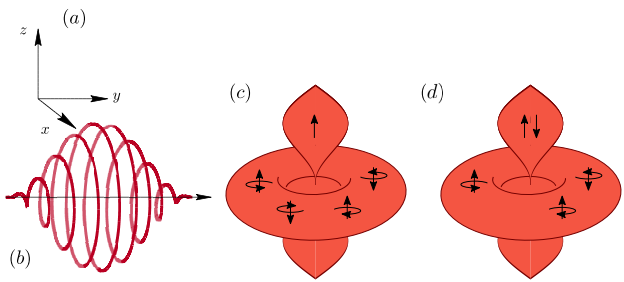}
    \caption{$(a)$ Laboratory fixed frame; $(b)$ vector potential, $\vec{A}(t)$, of a laser pulse circularly polarized in the  $xz$ plane; $(c)$ and $(d)$ sketch of the orbitals in the $2p$ shell of excited Ne$^*(1s^22s^2 2p^5[{}^2\text{P}^o_{1/2}]3s[^1\text{P}^o])$ with a hole in $(c)~2p_z$ (corresponding to $\mtot=0$) and $(d)~2p_{-1}$ (corresponding to $\mtot=1$). The vertical arrows in $(c)$ and $(d)$ denote the spin and the circular lines the magnetic quantum number of the electrons.}
    \label{fig:fig1}
\end{figure}

In the present work, we focus on the interaction of circularly polarized attosecond pulses with the first excited state in neon, the Ne$^*(1s^22s^2 2p^5[{}^2\text{P}^o_{1/2}]3s[^1\text{P}^o])$ state. We do, however, include a discussion of ionization by linearly polarized attosecond pulses, since this case, in combination with the results for circular polarization, allows us to discuss the polarization-dependence on the accessible ionization channels. We will use an approach that accounts for electron-electron correlation (the TD-RASSCF singles approach~\cite{Miyagi2014}). We imagine that the excited state is prepared by ultrashort-pulse excitation and we then probe it by an attosecond pulse, which initiates ionization. Here we investigate the interaction of aligned excited states, characterized by total magnetic quantum number $\mtot=0,\pm 1$, with an ultrashort XUV attosecond laser pulse polarized in the  $xz$-plane. The geometry is illustrated in Fig.~\ref{fig:fig1}. As is clear from the figure, we probe the atomic target with a pulse which for nonlinearly polarized light couples states with different azimuthal symmetry, allowing us to explore the electronic structure by means of the photoelectron momentum distribution (PMD). In this context, we find that some of the main peaks in the photoelectron spectra (PES) and PMD explicitly need correlation to appear, and we find that some of the peaks resulting from processes that do not need electron correlation to occur are suppressed compared to the former correlation-assisted peaks.
This paper is organized as follows: In Sec.~\ref{sec:theory_and_methods}, we describe the theoretical framework and the methodology. Next, in Sec.~\ref{sec:results}, we show the main results, including an analysis of the PES, PMD and the dipole moment, paying special attention to the asymmetries induced by the circularly polarized field. Finally, in Sec.~\ref{sec:conclusions}, we describe the conclusions of the present study and give an outlook including possible extensions. Throughout this work we use atomic units.

\section{Theory and methods}
\label{sec:theory_and_methods}
In this section we briefly describe the TD-RASSCF method and the computation of the PES and PMD.

\subsection{TD-RASSCF method}
\label{sec:TD-RASSCF_method}

\begin{figure}[b]
    \centering
    \includegraphics[width=.9\linewidth]{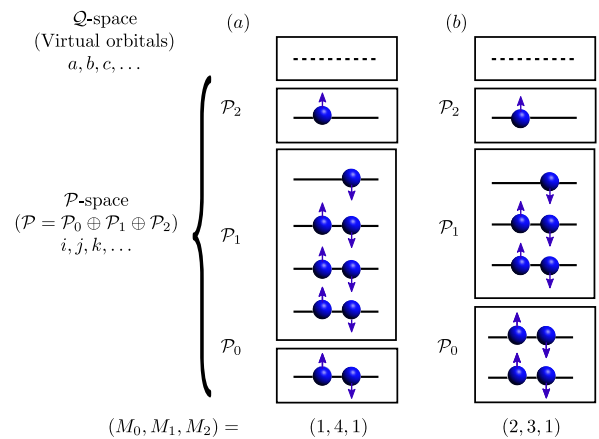}
    \caption{\label{fig:fig2} Restricted active spaces for neon corresponding to the partitions $(a)$ $(M_0,M_1,M_2)=(1,4,1)$ and $(b)$ $(2,3,1)$. Note that $(M_0,M_1,M_2)$ denote the number of spatial orbitals in the subspaces \pzero, \pone and \ptwo, respectively.}
\end{figure}

The TD-RASSCF method is used to propagate the  many-electron wave function. Since this methodology was explained in detail elsewhere~\cite{Miyagi2013, Miyagi2014,Miyagi2014b,Omiste2017_be, Omiste2018_neon,Madsen2018,Omiste2019}, the discussion here will be brief. In the TD-RASSCF approach~\cite{Miyagi2013,Miyagi2014b} the Ansatz of the many-body wave function reads
\begin{equation}
  \label{eq:wf_ansatz}
  \ket{\Psi(t)}=\sum_{\mathbf{I}\in\mathcal{V}} C_\mathbf{I}(t) \ket{\Phi_\mathbf{I}(t)},
\end{equation}
where the sum runs over the set of configurations $\mathcal{V}$, and $C_\mathbf{I}(t)$ and $\ket{\Phi_\mathbf{I}(t)}$ are the amplitudes and Slater determinants of the configuration $\mathbf{I}$, which are direct products of spin-up and spin-down strings, i.e., $\mathbf{I}=\mathbf{I_\uparrow}\otimes\mathbf{I_\downarrow}$, each of them including the indices of the spatial orbitals~\cite{Olsen1988,Klene2003}. Each Slater determinant is built from time-dependent spatial orbitals~$\{\ket{\phi_j(t)}\}_{j=1}^M$. In the case of MCTDHF, $\mathcal{V}\equiv \mathcal{V}_\textup{FCI}$, that is, the full configuration space~\cite{Koch2006}. In the case of TD-RASSCF~\cite{Miyagi2014b,Miyagi2013}, the configurations run in the restricted active space, $\mathcal{V}\equiv \mathcal{V}_\textup{RAS}$, which is defined by the restrictions on the excitations in the active space. The active orbital space $\mathcal{P}$ is divided into 3 subspaces:~$\mathcal{P}_0,\,\mathcal{P}_1$ and $\mathcal{P}_2$ [Fig.~\ref{fig:fig2}]. $\mathcal{P}_0$ constitutes the core, and its orbitals are fully occupied at all times. All the combinations of the orbitals in $\mathcal{P}_1$ are allowed and the number of occupied orbitals in $\mathcal{P}_2$ correspond to the permitted excitations from $\mathcal{P}_1$. In Fig.~\ref{fig:fig2}, the notation $(M_0, M_1, M_2)$ denotes the number of spatial orbitals in \pzero, \pone and \ptwo, respectively.

Here we apply TD-RASSCF including single (-S) excitations from the active space partition $\mathcal{P}_1$ to $\mathcal{P}_2$, with one and two orbitals in the $\mathcal{P}_0$ core space, as shown in Fig.~\ref{fig:fig2}.
 The TD-RASSCF theory is conveniently formulated in second quantization. We work in the spin-restricted framework, where a given configuration $\ket{\Phi_\mathbf{I}(t)}$, describing $N_e$ electrons, is constructed by $2M$ orbitals, formed by one of the $M$ spatial orbitals times the spin function. The Hamiltonian of a many-electron atom in the presence of a laser field reads 
\begin{equation}
  \label{eq:hamil_second}
H= \sum_{p,q} h^p_q(t)E^q_p +\frac{1}{2}\sum_{pqrs}v_{qs}^{pr}(t)E^{qs}_{pr},
\end{equation}
where the spin-free excitation operators $E^q_p$ and $E^{qs}_{pr}$ are defined as
\begin{equation}
  \label{eq:epq}
E^q_p=\sum\limits_{\sigma=\uparrow,\downarrow}b^\dagger_{p,\sigma} b_{q,\sigma},\quad E^{qs}_{pr}=\sum\limits_{\sigma=\uparrow,\downarrow}\sum\limits_{\gamma=\uparrow,\downarrow}b^\dagger_{p,\sigma}b^\dagger_{r,\gamma} b_{s,\sigma} b_{q,\gamma}
\end{equation}
with $b^\dagger_{p,\sigma}$ and $b_{p,\sigma}$ the creation and annihilation operators of a single spin-orbital $\ket{\phi_{p}(t)}\otimes\ket{\sigma}$. In Eq.~\eqref{eq:hamil_second}, the one-body, $h_q^p(t)$, and two-body, $v_{qs}^{pr}(t)$ matrix elements, are given by
\begin{eqnarray}
h_q^{p}(t)&=& \int\mathrm{d}\vec{r}\phi_p^\star(\vec{r},t)h(\vec{r},t)\phi_q(\vec{r},t),\\
\label{eq:vqspr}
v_{qs}^{pr}(t)&=&  \int\int\mathrm{d}\vec{r}\mathrm{d}\vec{r}'\cfrac{\phi_p^\star(\vec{r},t)\phi_r^\star(\vec{r}',t)\phi_q(\vec{r},t)\phi_s(\vec{r}',t)}{|\vec{r}-\vec{r}'|},
\end{eqnarray}
where the one-body operator in the dipole approximation is
\begin{equation}
\label{eq:h_1st_quantization}
  h(\vec{r},t)= \cfrac{p^2}{2}-\cfrac{Z}{r}+  \vec{E}(t)\cdot \vec{r},
\end{equation}
where $\vec{E}(t)$ is the electric field of the laser pulse.

 The equations of motion (EOM) determining the time evolution of the amplitudes $C_\mathbf{I}(t)$ and the orbitals $\ket{\phi_i(t)}$ were given in Refs.~\cite{Miyagi2014b,Miyagi2013} (see also Refs.~\cite{Lode2020,Omiste2019,Omiste2017_be,Omiste2018_neon}) and will not be reproduced here. They are obtained from the Antsatz~\eqref{eq:wf_ansatz} and the Dirac-Frenkel-McLachlan TD variational principle. Our implementation benefits from the properties of the finite-element discrete-variable representation (FE-DVR) for the radial part~\cite{Omiste2017_be} and the coupled basis method to speed up the computation of the angular part~\cite{Omiste2017_be,Omiste2018_neon}. The PES and PMD were calculated as described in Refs.~\cite{Omiste2017_be,Madsen2018}, and we followed that description, projecting on Coulomb waves in the asymptotic region.

\section{Results}
\label{sec:results}

We explore the laser-induced ultrafast ionization of the first excited state of neon, Ne$^*(1s^22s^22p^5[^{2}P^o_{1/2}]3s[^1\text{P}^o])$ (from now on Ne$^*$) is 16.848~eV above the ground state~\cite{NIST_ASD_2019}. We imagine that this state is prepared by ultrafast pulse excitation. The XUV pulse that probes this excited state is described by the vector potential
\begin{equation}
    \label{eq:vector_potential}
    \vec{A}=\frac{A_0}{\sqrt{1+\varepsilon^2}}\cos^2[\omega t/(2 n_p)]\left(
    \begin{array}{c}
         \varepsilon\cos(\omega t+\varphi) \\
         0 \\
         \sin(\omega t+\varphi)\\
    \end{array}
    \right),
\end{equation}
where $A_0$ corresponds to a maximum intensity of  $\SI{e14}{\intensity}$, $\varepsilon$ is the ellipticity, $n_p$ the number of cycles and $\varphi$ denotes the carrier envelope phase (CEP). We use $\varphi=0$ unless otherwise stated. We consider linearly ($\varepsilon=0$) and counter-clock circularly polarized ($\varepsilon=1$) pulses. We consider $n_p=5, 7, 10$ and angular frequencies of $\omega = 50, 70$~and $105$~eV, corresponding to pulses with durations between $393.8$ and $413.6$ attoseconds. Even though the intensity is as high as 10$^{14}$ W/cm$^2$, the dynamics is still mainly in the perturbative regime for the laser-atom interaction. This can, e.g., be seen by considering the magnitude of the ponderomotive energy $U_p$, which always fulfils $U_p/\omega \ll 1$. The electric field of the laser pulse enters Eq.~\eqref{eq:h_1st_quantization} and is obtained from the vector potential in Eq.~\eqref{eq:vector_potential} as $\vec{E}(t)=-\partial_t\vec{A}(t)$. This way of specifying the field, ensures that $\vec{E}(t)$ contains no unphysical DC components~\cite{Madsen2002}.

\subsection{Initial state}
\label{sec:initial_state}

Self-consistent TD methods normally apply imaginary-time propagation to obtain the absolute ground state. Here, we apply a similar method to obtain the excited initial state Ne$^*$ by restricting the symmetry of the state used for propagation in imaginary time. Note that recently, a new method was proposed to compute excited states using an extension of MCTDHF that generalizes the state-average approach of CASSCF methods~\cite{Lotstedt2020}. Specifically, in our present case, the first excited state of Ne corresponds to the excitation of a $2p$ electron to the $3s$ shell,~\ie, the main configuration of the excited state is ($1s^22s^22p^5[\doubletpodd]3s[^1P^o]$). This state is the lowest of ${}^{1}\text{P}^o$ symmetry. Thus, the imaginary-time propagation of a guess function belonging to this symmetry and having a non-vanishing overlap with the ${}^{1}\text{P}^o$-state sought for, converges to it. We use two different RAS schemes: $(M_0,M_1,M_2)=(1,4,1)$ and $(2,3,1)$ including only single excitations from $\mathcal{P}_1$ to $\mathcal{P}_2$ (see Fig.~\ref{fig:fig2}).  The orbitals contributing to the initial guess functions are chosen as hydrogenic $1s,2s,2p$ and $3s$ orbitals for $Z=10$. We impose the guess function to be odd under inversion by setting to zero the amplitudes of the configurations with $\sum_k \ell_k$ even, with $\ell_k$ the orbital angular momentum of the $k$th orbital. In addition, since Ne$^*$ has total orbital angular momentum $L=1$, we can obtain the states corresponding to the total magnetic quantum number $M_\text{tot}=-1,0$ and $1$ by setting the amplitudes of the configuration with \mtot~different from these three values to zero. We benefit from the TD-RASSCF method, since it preserves $M_\text{tot}$ and the magnetic quantum number of each orbital during the propagation if the rotation around the  $z$-axis is a symmetry operation~\cite{Omiste2019}, as is the case for a field-free atom.

\begin{figure}
    \centering
    \includegraphics[width=\linewidth]{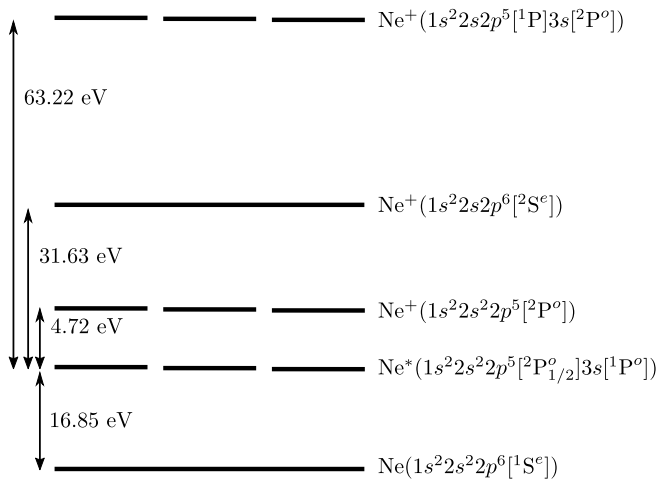}
    \caption{\label{fig:fig3} Excitation energies and ionization potential for the states involved. The ionization threshold for Ne$^+(1s^22s2p^5[^1\text{P}]3s[^2\text{P}^o])$ is computed using OpenMolcas~\cite{Aquilante2016}, see text. The rest are taken from Ref.~\cite{NIST_ASD_2019}. The broken lines symbolize the degeneracy of the states with respect to \mtot.}
\end{figure}

Let us highlight that for the conditions imposed on the orbitals and configurations, the same configurations contribute to the wave function of the excited states with a given $M_\text{tot}$ for both the RAS considered. Therefore, we obtain the same energy for the excited initial state with both RAS. However, the ground state energy depends on the RAS since different configurations contribute to its wave function. Specifically,  the excited state energy obtained is $-127.994$~a.u. for all $M_\text{tot}$, which corresponds to an excitation energy of $0.55398$~a.u=$15.068$~eV for $(M_0,M_1,M_2)=(2,3,1)$, and $0.56718$~a.u.=$15.428$~eV for $(M_0,M_1,M_2)=(1,4,1)$ with respect to the corresponding ground state energy. As we illustrate in Fig.~\ref{fig:fig3}, the data-base value $16.848$~eV~\cite{NIST_ASD_2019}, differs from our computation because the eigenstates are obtained with different correlation as the number of configurations included in each RAS is different~\cite{Helgaker2000}. We do not expect that this shift in the energy affects the shape of the PES. Shifts in energy are typical in many-electron studies of photoionization in atoms and molecules and reflect the quality of the structure calculation. In the next section, we check the accuracy of the TD-RASSCF by determining the ionization thresholds by means of the PES.

\subsection{Linearly polarized laser field}
\label{sec:linearly_polarized_laser_field}
In this section we~ explore the effects of the anisotropy of Ne$^*$ by analyzing the PES induced by a linearly polarized XUV attosecond laser pulse. The linear polarization of the laser is chosen parallel to the $z$-axis,~\ie, parallel to the quantization axis of the excited states, to preserve the azimuthal symmetry of the system (see Fig.~\ref{fig:fig1}). Investigation of the physics related to a variation of the angle between the quantization axis of the excited state and the direction of the linear polarization is left for a future study.

The one-photon ionization channels of Ne$^*$ differ for the different values of \mtot. The ionization reaction is given by
\begin{equation}
    \label{eq:ionization_reaction_general}
\text{Ne}^*(1s^22s^22p^5[\doubletpodd_{1/2}]3s[^1P^o])[\mtot]+\gamma\rightarrow \text{Ne}^++e^-(\ell m),  
\end{equation}
where $\mtot=0,\pm 1$ and the pair $(\ell m)$ describe the orbital angular momentum and magnetic quantum number of the outgoing electron, respectively. Let us emphasize that the angular momentum of the outgoing electron is restricted due to the symmetry of the parent ion defining the ionization channel. Specifically, for one-photon absorption, the difference in total angular momentum of the final ionization state (including the parent ion and the outgoing electron) and the initial state fulfills $|\Delta L|=0,\,1$, $\mtot=\mtot^\prime+m$, with~$\mtot^\prime$ the total magnetic quantum number of the parent ion, and the initial and final states have opposite parities. The allowed quantum numbers for the different ionization channels are collected in Table~\ref{tab:ionization_linear}.
\begin{table}[h]
    \centering
    \begin{ruledtabular}
    \begin{tabular}{c|c|c}
        Parent ion \& &  \mtot  & $(\ell,m)$ of outgoing  \\
        ionization potential &   &  electron\\
        from Ne$^*$ &   &  \\
        \hline
        $\text{Ne}^+ 
(1s^22s^22p^5[\doubletpodd])$ & -1 & (1, -1); (1, 0); (3, -1); (3, 0)\\
        $4.72$~eV & 0 & $(1, \pm 1)$; $(1, 0)$; $(3, \pm 1)$; $(3, 0)$\\
         & 1 & (1, 1); (1, 0); (3, 1); (3, 0)\\
         \hline
         $\text{Ne}^+ (1s^22s2p^6[\doubletseven])$& -1 & $(2,-1)$\\
         $31.63$~eV & 0 & $(0,0)$; $(2,0)$\\
         & 1 & (2,1)\\
         \hline
        $\text{Ne}^+ (1s^22s2p^5[\singletp]3s[\doubletpodd])$ & -1 & (1, -1); (1, 0); (3, -1); (3, 0)\\
         $63.22$~eV & 0 & $(1, \pm 1)$; $(1, 0)$; $(3, \pm 1)$; $(3, 0)$\\
         & 1 & (1, 1); (1, 0); (3, 1); (3, 0)\\
    \end{tabular}
    \end{ruledtabular}
    \caption{\label{tab:ionization_linear} Single-photon ionization channels after the interaction of Ne$^*$ with a pulse, linearly polarized along the $z$-axis [see Eq.~\eqref{eq:ionization_reaction_general}]. The ionization potentials are given with respect to the ground state, and are obtained as described in the caption of Fig.~\ref{fig:fig3}.}
\end{table}
The most remarkable feature is that after the ionization for $\mtot=\pm 1$ into the $\text{Ne}^+ (1s^22s2p^6[\doubletseven])$ channel, the outgoing electron cannot have $l=0$, due to the symmetry restrictions. However, for $\mtot=0$, $\ell=0$ is allowed and this difference has an impact on the PES as we describe below.
\begin{figure}
    \centering
    \includegraphics[width=\linewidth]{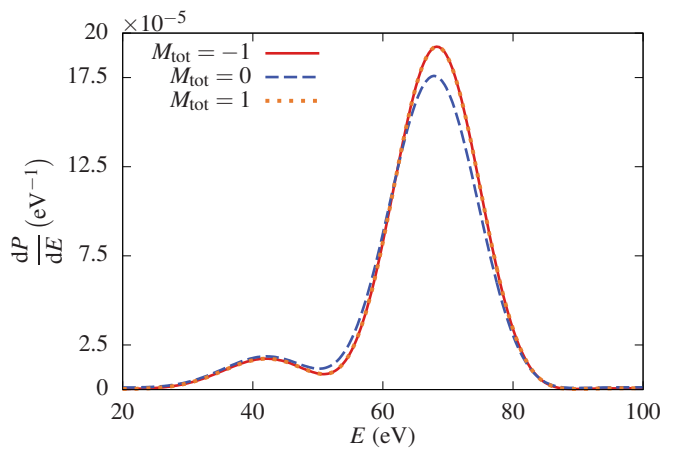}
    \caption{\label{fig:fig4} Photoelectron spectrum of excited  Ne$^*(1s^22s^22p^5[\doubletpodd_{1/2}]3s[^1P^o])[\mtot]$ after interacting with a linearly polarized laser pulse with a central photon energy of $105$~eV, a duration of 10 cycles and $\varphi=0$ [see Eq.~\eqref{eq:vector_potential}]. The RAS chosen is $(M_0,M_1,M_2)=(1,4,1)$, whose results are indistinguishable from those of $(M_0,M_1,M_2)=(2,3,1)$ on the scale of the figure.}
\end{figure}
 
Let us now describe the PES in light of these considerations. In Fig.~\ref{fig:fig4} we show the PES following ionization of Ne$^*$ with $\mtot=0,\pm 1$ by a linearly polarized laser pulse with a central photon energy of $105$~eV and $10$ cycles. First of all, we emphasize that the PES for the RAS considered in this work,~\ie, $(M_0,M_1,M_2)=(2,3,1)$ and $(1,4,1)$, are indistinguishable on the scale of the figure, which manifests the convergence of the results and highlights one of the advantages of the TD-RASSCF method,~\ie, the possibility to work with different active spaces and restrictions, while keeping the number of configurations manageable. We observe two peaks in both PES: the highest one is at around $68$~eV, and corresponds to ionization into final state $\text{Ne}^+ (1s^22s2p^6[\doubletseven])$. According to the data values for the  ionization threshold from Ne$^*$~\cite{NIST_ASD_2019}, the peak should be located at $73.369$~eV. The secondary peak is at around $40$~eV and the PES for $\mtot=0,\pm 1$ overlap in this energy range. By a calculation with  OpenMolcas~\cite{Aquilante2016} using a cc-ptvz basis, we have identified this second peak to stem  from ionization into the parent ion $\text{Ne}^+(1s^22s2p^5[\singletp]3s[\doubletpodd])$, which is in good agreement with the literature~\cite{Feist2014}. The ionization yield corresponding to the parent ion $\text{Ne}^+ (1s^22s2p^6[\doubletseven])$ differs for $|\mtot|=1$ and $\mtot=0$, because for the latter case, the ejected electron can acquire a vanishing angular momentum~[see Table~\ref{tab:ionization_linear}]. This channel has a different cross section than the pathway followed if the ionized electron is characterized by $\ell=2$. Let us also remark that the peak corresponding to the parent ion $\text{Ne}^+ (1s^22s^22p^5[\doubletpodd$]), expected at $\sim 100$~eV for a photon energy of $105$~eV, is highly suppressed, since the ratio of the ionization potential ($4.94$~eV) over the photon energy is $0.047\ll 1$, and the photoionization cross section decreases with the continuum electron energy. This effect has also been observed in computation for the photoionization of Magnesium~\cite{Kochur2001}.

\begin{figure}
    \centering
    \includegraphics[width=\linewidth]{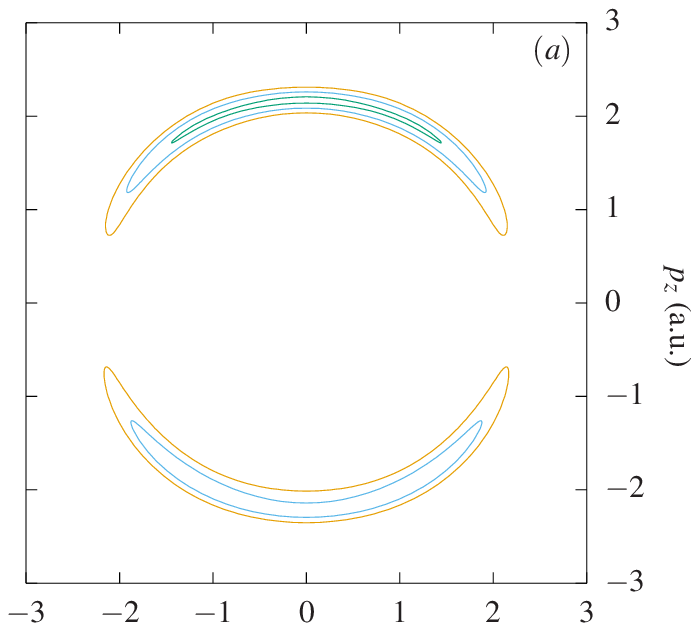}
    \includegraphics[width=\linewidth]{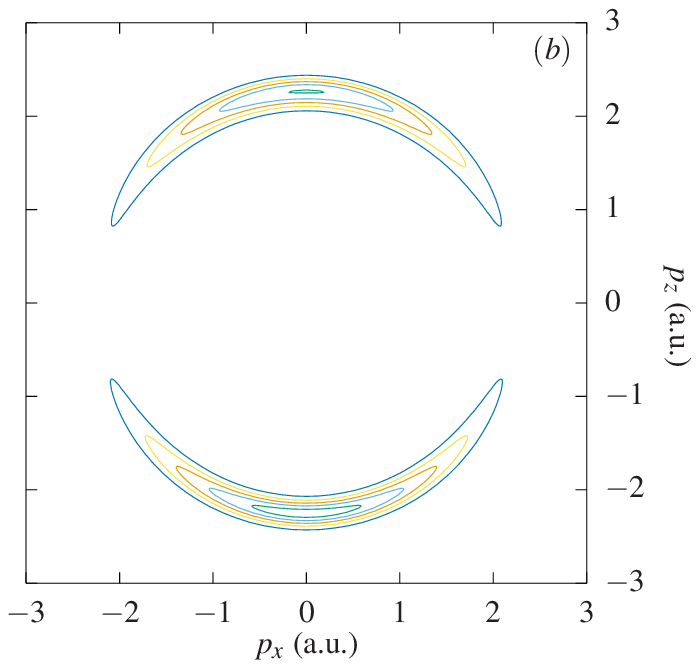}
    \caption{\label{fig:fig5} PMD for a linearly polarized XUV pulse with $\omega=105$~eV and 10 cycles and $\varphi=0$~[see Eq.~\eqref{eq:vector_potential}], for $p_y=0$ for the excited states of Ne$^*$ with (a) $M_\text{tot}$= 0 and (b) $\pm 1$. The polarization of the laser pulse is in the  $z$-direction.}
\end{figure}

Next, we compare the PMD of the ejected electron for initial $\mtot = 0$ and $\pm 1$ at $p_y = 0$, shown in Fig.~\ref{fig:fig5}. We observe that the PMDs peak along the polarization direction of the linearly polarized pulse and that the distribution for $\mtot=0$ is wider than for $|\mtot|=1$. We may associate this difference with interference between outgoing $s$- and $d$-waves, which for the dominating Ne$^+(1s^22s2p^6[^2\text{S}^e])$ channel is present in the \mtot=0 case, but absent for $|\mtot|=1$, since only outgoing $d$-waves contribute to the PMD in the latter case [see Table~\ref{tab:ionization_linear}].

Finally, one remarkable feature of the PES is that the highest peak in Fig.~\ref{fig:fig4} implies the rearrangement of the two electrons, and can be understood in terms of two processes illustrated in Figs.~\ref{fig:fig6}
 and~\ref{fig:fig7}: 
\begin{figure}
    \centering
    \includegraphics[width=.9\linewidth]{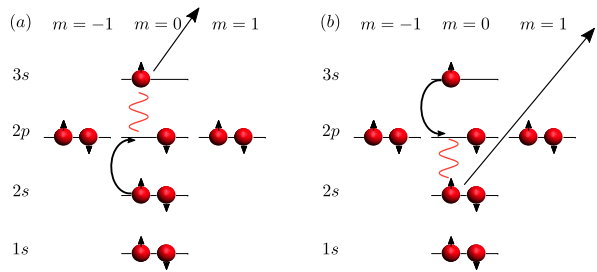}
    \caption{For $\mtot=0$, electronic excitation mediated by the laser field which leads to ionization into the channel $\text{Ne}^+ (1s^22s2p^6[\doubletseven])+e^-(\ell m)$ from Ne$^*$. The two competing ionization processes are $(a)$ the excitation of a $2s$ electron to the $2p$ subshell and subsequent ionization of the $3s$ electron and $(b)$ the de-excitation of the $3s$ electron to the $2p$ subshell, leading to the ionization of one of the electrons in the $2s$ subshell.}
    \label{fig:fig6}
\end{figure}
(i) the excitation of a $2s$ electron to the $2p$ subshell and the ionization of the $3s$ electron or (ii) the de-excitation of the $3s$ electron to the $2p$ subshell and the ionization of one of the electrons in the $2s$ subshell. Note that these pathways cannot be mediated by a one-photon process without including electron correlation as done with the present TD-RASSCF-S method. Also note that the proposed pathways in panels (a) and (b) of each figure contribute simultaneously to the ionization process and their relative contribution cannot be straightforwardly distinguished. We note that the ratio of the signal into the correlation-free peak at $\sim$ 100 eV for the Ne$^+(1s^22s^22p^5[^2\text{P}^o])$ channel to the signal into the correlation-assisted peak at $68$~eV for the Ne$^+(1s^22s2p^6[^2\text{S}^e])$ channel is much smaller than unity, which reflects the decrease in the photoionization cross section with photoelectron energy. The smallness of the ratio indicates that correlation is not only crucial for making the peak at $68$~eV accessible, it also enhances the yield into that channel. In other words, the results show that in the final channel a very asymmetric, unequal energy-sharing between excitation in the ion and energy in the continuum is less favorable than a more symmetric energy sharing between excitation in the ion and energy in the continuum, even when these latter channels need correlation to be accessible.  
\begin{figure}
    \centering
    \includegraphics[width=.9\linewidth]{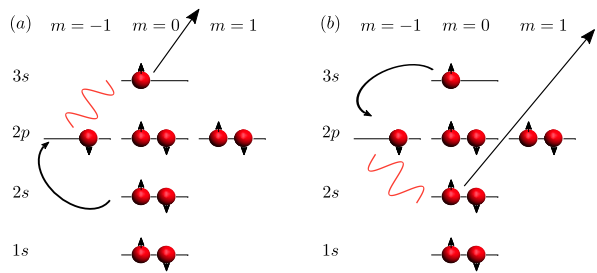}
    \caption{For $\mtot=1$, electronic excitation mediated by the laser field which leads to ionization into the channel $\text{Ne}^+ (1s^22s2p^6[\doubletseven])+e^-(\ell m)$ from Ne$^*$. The two competing ionization processes are $(a)$ the excitation of a $2s$ electron to the $2p$ subshell and subsequent ionization of the $3s$ electron and $(b)$ the de-excitation of the $3s$ electron to the $2p$ subshell, leading to the ionization of one of the electrons in the $2s$ subshell. The $\mtot=-1$ is similar except that the $2p_{1}$ orbital is involved.}
    \label{fig:fig7}
\end{figure}

\subsection{Circularly polarized laser field}
\label{sec:circularly_polarized_laser_field}
In this section we compute the interaction of Ne$^*$ with circularly polarized XUV attosecond laser pulses propagating along the  $y$-axis,~\ie, polarized in the  $xz$ plane (see Fig.~\ref{fig:fig1}). We also consider variation of the dipole moment during the interaction. We use a positive ellipticity ($\varepsilon=1$),~\ie, the counter-clockwise rotating vector potential in Eq.~\eqref{eq:vector_potential} for all the computations.

\subsubsection{Photoelectron spectrum asymmetry}
\label{sec:pes_asymmetry}

First, we analyze the PES after the interaction with the laser. The PES probes the electronic correlation and structure of Ne$^*$. In Table~\ref{tab:ionization_circular_xz} we collect the channels accessible following ionization by a circularly polarized pulse in the  $xz$-plane (see Fig.~\ref{fig:fig1}).

\begin{table}[h]
    \centering
    \begin{ruledtabular}
    \begin{tabular}{c|c|c}
              Parent ion \& &  \mtot  & $(\ell,m)$ of outgoing  \\
        ionization potential &  &  electron\\
        from Ne$^*$ &   &  \\
        \hline
        $\text{Ne}^+ 
(1s^22s^22p^5[\doubletpodd])$ & -1 & $(1, m^\prime)$; $(3,-2)$; $(3, \pm 1)$; $(3, 0)$\\
        4.94~eV & 0 & $(1, m^\prime)$; $(3, \pm 1)$; $(3, 0)$\\
         & 1 & $(1, m^\prime)$; $(3,2)$; $(3, \pm 1)$; $(3, 0)$\\
         \hline
         $\text{Ne}^+ (1s^22s2p^6[\doubletseven])$& -1 & $(0,0)$; $(2,-2)$; $(2,-1)$; $(2,0)$\\
        31.85~eV & 0 & $(0,0)$; $(2,0)$, $(2,\pm 1)$\\
         & 1 & $(0,0)$; $(2,0)$; $(2,1)$; $(2,2)$\\
         \hline
        $\text{Ne}^+ (1s^22s2p^5[\singletp]3s[\doubletpodd])$ & -1 & $(1, m^\prime)$; $(3,-2)$; $(3, \pm 1)$; $(3, 0)$\\
        63.44~eV & 0 & $(1, m^\prime)$; $(3, \pm 1)$; $(3, 0)$\\
         & 1 & $(1, m^\prime)$; $(3,2)$; $(3, \pm 1)$; $(3, 0)$\\
    \end{tabular}
    \end{ruledtabular}
    \caption{\label{tab:ionization_circular_xz} Single photon absorption ionization channels the interaction of Ne$^*$ with a pulse circularly polarized in the  $xz$-plane (see Fig.~\ref{fig:fig1}). In the notation $(\ell, m^\prime)$, $m^\prime$ can take any value from $-\ell$ to $\ell$. The ionization potentials are as in Table~\ref{tab:ionization_linear}.}
\end{table}

\begin{figure}
    \centering
    \includegraphics[width=\linewidth]{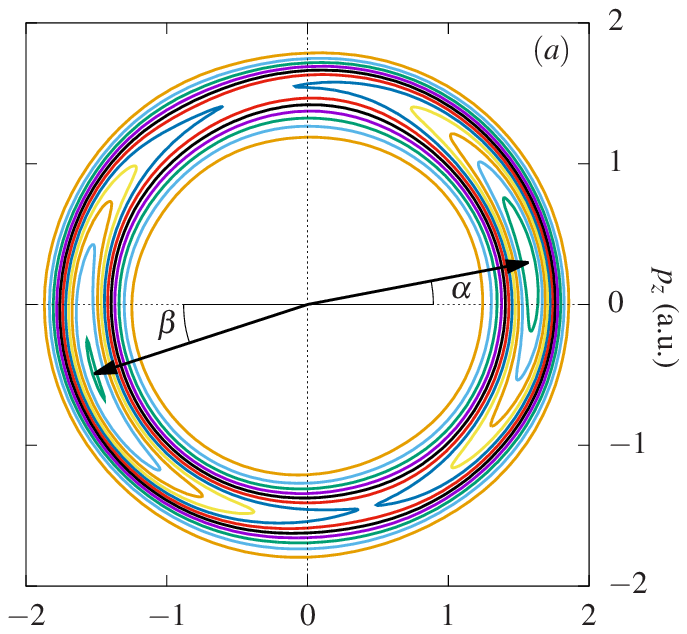}
    \includegraphics[width=\linewidth]{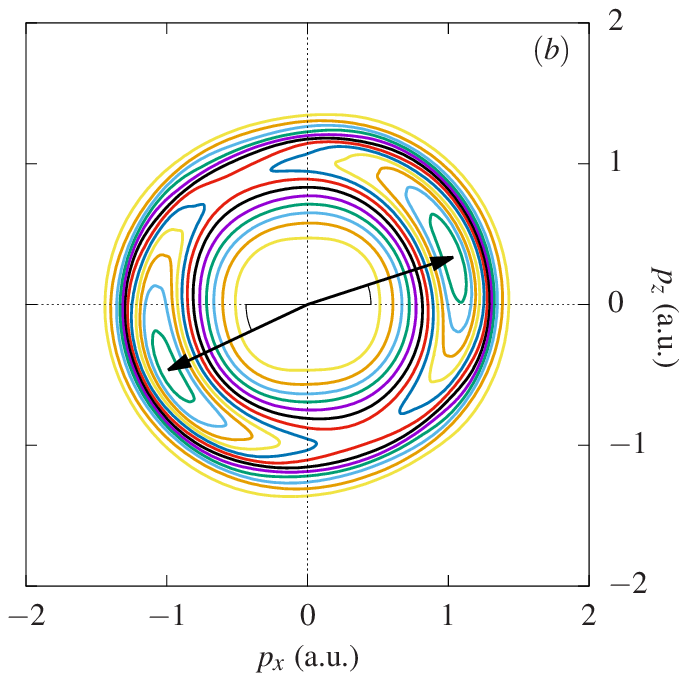}
    \caption{\label{fig:fig8} PMD in the polarization plane emitted by Ne$^*$ and $M_\text{tot}= 0$ at $p_y=0$ after interacting with a circularly polarized XUV laser with $\varphi=0$ and $(a)$ $70$~eV and 7 cycles and $(b)$ $50$~eV and 5 cycles. The arrows point to the maxima of the PMD. These results are obtained with RAS $(M_0,M_1,M_2)=(1,4,1)$ with single excitations.}
\end{figure}

\begin{figure}
    \centering
    \includegraphics[width=\linewidth]{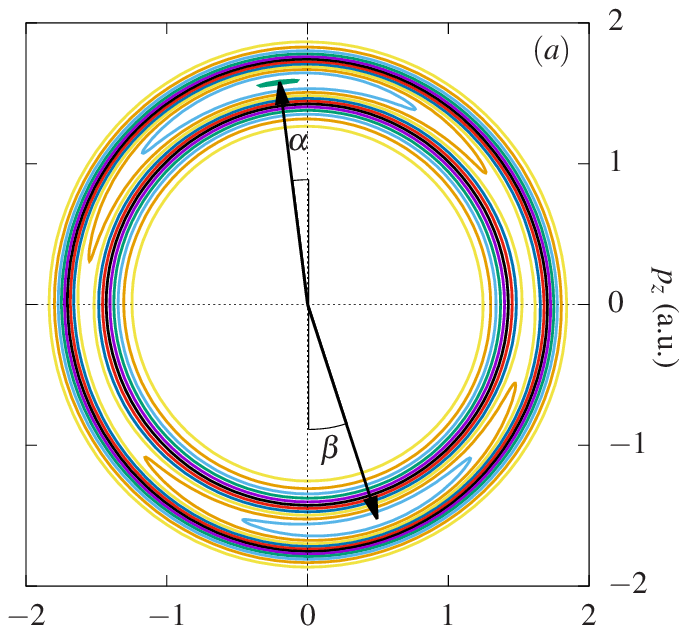}
    \includegraphics[width=\linewidth]{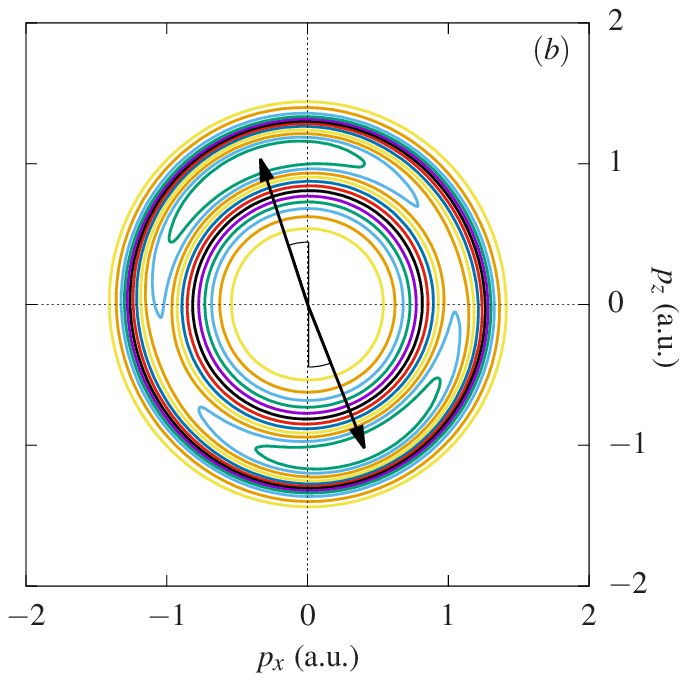}
    \caption{\label{fig:fig9} PMD in the polarization plane emitted by Ne$^*$ and $M_\text{tot}= 1$ at $p_y=0$ after interacting with a circularly polarized XUV laser with $\varphi=0$ and $(a)$ $70$~eV and 7 cycles and $(b)$ $50$~eV and 5 cycles. The arrows point to the maxima of the PMD. These results are obtained with RAS $(M_0,M_1,M_2)=(1,4,1)$ with single excitations.}
\end{figure}

We focus on correlation-assisted ionization leaving the parent ion in the state Ne$^+(1s^22s2p^6[\doubletseven])$. The cross section for this channel is larger compared with those of the other channels as was also the case for linearly polarized light. Let us again remark that this fact is counter-intuitive since the ionization into this channels needs electron correlation, as we discussed in Sec.~\ref{sec:linearly_polarized_laser_field} and illustrated in Figs.~\ref{fig:fig6} and~\ref{fig:fig7}. In Figs.~\ref{fig:fig8} and~\ref{fig:fig9} we show the PES for $p_y=0$ for $\mtot=0,1$ and two XUV pulses: (i) $\omega=70$~eV and $7$ cycles and (ii) $\omega=50$~eV and $5$ cycles. We observe that the PMDs in the polarization plane for $\mtot=0$ and $1$ are slightly tilted with respect to the  $x$ and  $z$ axis, respectively. Note that the PMDs for $\mtot=-1$ is the same than for $\mtot=1$ for the alignment considered.

In order to understand the PMDs, we analyze the orbitals of the $2p$ shell, shown in Fig.~\ref{fig:fig1}. For instance, the case of Ne$^*[\mtot=0]$ corresponds to exciting a $2p_z$ electron to the $3s$ shell leaving a hole in the $2p_z$ orbital [see Fig.~\ref{fig:fig1}(c)]. The hole in the $2p_z$ orbital implies that there is an excess of electronic charge around the $xy$-plane, being more polarizable under the impact of the laser around this plane. Therefore, we would expect that the emitted electron is ejected along the $x$ axis, given that the laser is polarized in the $xz$-plane. However, the spectrum is tilted with respect to this axis, as shown in Fig.~\ref{fig:fig8}. On the other hand, if the hole is in $2p_{\pm 1}$ orbitals, then the defect of electronic charge is located around the $xy$ plane, and therefore the excess of charge is along the  $z$-axis, as sketched in Fig.~\ref{fig:fig1}(d). This would imply the maximum of the PMD to be located along the  $z$ axis, but we find that it is tilted with respect to that axis. The inclination of the maximum is counter-clock wise, which corresponds to the direction of rotation of the polarization axis of the laser. We note in passing that a shift of the maximum of the distribution compared to that expected was also observed for strong fields at $\sim$800~nm-and also in the direction of rotation of light~\cite{Eckle2008a,Eckle2008b,Pfeiffer2012,Martiny2009}. In order to measure the rotation of the PMDs in the polarization plane, we denote by $\alpha$ the angle formed by the line linking the maximum for $p_z>0$ with the origin and the $x$ ($z$)-axis for $\mtot=0$ ($\mtot=1$). To illustrate the asymmetry we also introduce $\beta$, which is the angle formed by the line linking the maximum for $p_z<0$ with the origin and the $x$ ($z$)-axis for $\mtot=0$ ($\mtot=1$) [see Fig.~\ref{fig:fig8}]. In Table~\ref{tab:rotation_pes} we collect $\alpha$ and $\beta$ for some laser field parameters and initial states.
\begin{table}[h]
    \centering
    \begin{ruledtabular}
    \begin{tabular}{|c|c|c|c|c|}
        \mtot & $\omega$~(eV) & number of cycles & $\alpha~(\degree)$ & $\beta~(\degree)$ \\
        \hline
        0& 50 & 5 & 18 & 25\\
         0& 70 & 5 & 11 & 18 \\
0& 70 & 7 & 11 & 18\\
0& 70 & 9 & 11 & 18\\
\hline
1& 50 & 5 & 18 & 21.6 \\
1& 70 & 7 & 7.2 & 18 \\
    \end{tabular}
    \end{ruledtabular}
    \caption{\label{tab:rotation_pes} Counter-clockwise rotation angles of the PMD in the polarization plane corresponding to Ne$^*(1s^22s^22p^5[\doubletpodd_{1/2}]3s[^1P^o])[\mtot]$. Note that $\alpha$ and $\beta$ are the angles with respect to the  $x$-axis for $\mtot=0$ and the  $z$-axis for $\mtot=1$, respectively. See Figs.~\ref{fig:fig8}(a) and~\ref{fig:fig9}(a), respectively.}
\end{table}
For $\omega=70$~eV we find that $\alpha=11\degree$ and $\beta=18\degree$ for $n_p=5,7$ and $9$, thus, the rotation of the PMD is independent of the duration of the XUV pulse in this regime. This process can be understood as the interference of channels corresponding to the ejected electron along the $z$ direction (for example, $p_z$-like wave function) and the $x$ direction ($p_x$-like wave function). That is to say, the angles $\alpha$ and $\beta$ are closely related to the atomic target as well as to the cross section of each ionization channel. Besides, the rotation of the PMDs is different for $\mtot=0$ and $|\mtot|= 1$, which is explained by the different electronic distributions, as shown in Fig.~\ref{fig:fig1} and the different allowed ionization channels.

To rationalize the asymmetry of the PMD, we express the  outgoing electron wave function in its spherical harmonics content. As an example, we consider the case $\mtot=0$. Here the angular part corresponding to the dominant channel can be expressed as
\begin{eqnarray}
\nonumber
\psi(\Omega)&\sim& Y_{00}(\Omega)+\gamma_0 Y_{20}(\Omega)+\gamma_{-1}Y_{2-1}(\Omega)- \gamma_1 Y_{21}(\Omega),\\
\label{eq:psi_circular_m0}
\end{eqnarray}
where $\gamma_0,~\gamma_{\pm 1}\in\mathds{C}$ and include information about the relative amplitudes of the different angular parts. 
The expression~\eqref{eq:psi_circular_m0} implies that the PMD in the polarization plane is invariant under a two-fold rotation around the axis perpendicular to the  $xz$-plane,~\ie, the   $y$-axis, and therefore, $\gamma_{-1}=\gamma_1$. However, the PMD in the polarization plane are slightly asymmetric, that is to say, $\gamma_{-1}\ne\gamma_1$~\cite{Omiste2011a,Zare1988} and similarly for terms with higher values of odd $\ell$. Thus, this asymmetry may be explained by two different sources. On the one hand, considering the short duration and intensity of the laser, which leads to small contributions beyond the first-order perturbation theory limit, and then, it depends on the CEP. On the other hand, the electronic correlation and interference of different ionization channels can also break the two-fold symmetry. A deeper analysis on the nature of the shifts in the ejection angles is needed to understand the underlying mechanism, which may be related to the shifts found in the strong field regime~\cite{Martiny2009} or the directional emission investigated in Ref.~\cite{Ivanov2013}.

\subsubsection{Dipole moment}
\label{sec:dipole_moment}
\begin{figure}
    \centering
    \includegraphics[width=\linewidth]{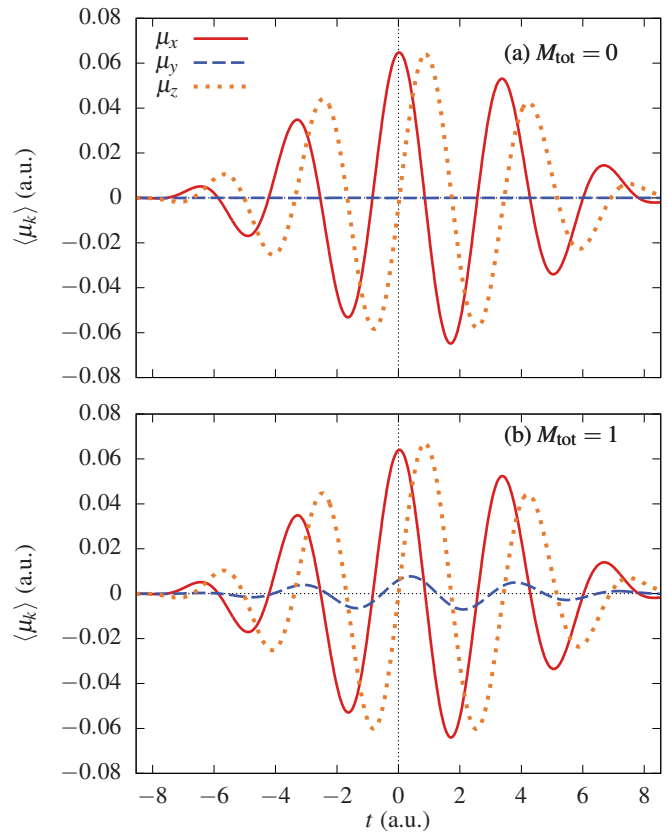}
    \caption{\label{fig:fig10} Dipole moment along the  $x$,  $y$ and   $z$ axis for Ne$^*$ interacting with a circularly polarized XUV laser pulse with a central photon energy of $50$~eV and a duration of 5 cycles, for (a) $\mtot=0$ and (b) $\mtot=1$. These results are obtained with RAS $(M_0,M_1,M_2)=(1,4,1)$ with single excitations.}
\end{figure}

Next, in order to understand the effect of the laser on the electronic cloud, we show the dipole moment,~$\expected{\vec{\mu}}(t)=-\melement{\Psi(t)}{\vec{r}}{\Psi(t)}$, during the propagation of the laser for $\mtot=0$ and $1$ for a central photon energy of $50$~eV and $n_\text{cycles}=5$ in Fig.~\ref{fig:fig10}. The trend is similar for $\omega =70$ and 105 eV and $n_p = 7$ and 10, respectively (not shown).

The dynamics of the dipole moment are very similar for $\mtot=0$ and $1$, since the pulse mainly affects the electron in the outer shell,~\ie, the $3s$ electron, which is common for both states. However, there are some small differences in the  $xz$-plane due to the different charge distribution of the electrons in the $2p$ subshell. This shell presents a different angular distribution for each state, therefore, the dynamical polarization is not the same. For example, the highest peak of $\mu_z$ is slightly larger for \mtot=1 than for \mtot=0. 

On the other hand, Fig.~\ref{fig:fig10} shows that the circularly polarized light is able to probe the chirality of the system. Specifically, for $\mtot=1$ the dipole moment along the propagation axis of the pulse,~\ie, $\mu_y$ is not zero. Moreover, if we set $\mtot=-1$ as initial state, we observe that $\mu_y$ changes sign whereas $\mu_x$ and $\mu_z$ show the same dynamics, that is to say, Ne$^*$ with $\mtot=\pm 1$ can be distinguished using circularly polarized light.
On the contrary, for $\mtot=0$, $\mu_y$ is zero (smaller than $10^{-14}$ in our computation). To understand these effects we turn to the symmetrical properties of the Ne$^*$. If the electronic wave function is invariant under reflections in the  $xz$ plane $\sigma_{xz}: \theta\rightarrow \theta,\phi\rightarrow 2\pi-\phi$, $\mu_y$ is zero during the laser pulse propagation. Note that $\theta$ and $\phi$ are the usual polar coordinates. However, if the wave function is not invariant under $\sigma_{xz}$ the laser can couple states with different orientations along the  $y$-axis, leading to net orientation. Applying $\sigma_{xz}$ to a spherical harmonic we obtain~\cite{Omiste2011a,Zare1988}
\begin{equation}
\label{eq:sigma_yjm}
    \sigma_{xz}Y_{\ell,m}(\theta,\phi)=(-1)^mY_{\ell,-m}(\theta,\phi),
\end{equation}
therefore, the orbitals map as $\sigma_{xz} s = s$ and $\sigma_{xz} p_m =(-1)^m p_{-m}$. Thus, since the main configuration of Ne$^*(\mtot=0)$ corresponds to $1s^22s2p_{-1}^22p_{+1}^22p_03s$, it is easy to show that it is even under $\sigma_{xz}$. However, the main configuration for Ne$^*[\mtot=1]$ is $1s^22s2p_{-1}^22p_{0}^22p_{+1}3s$, then 
\begin{equation}
    \sigma_{xz}(1s^22s2p_{-1}^22p_{0}^22p_{+1}3s)=(-1)(1s^22s2p_{+1}^22p_{0}^22p_{-1}3s).
\end{equation}
As we see, under the application of $\sigma_{xz}$ on Ne$^*[\mtot=1]$ we get Ne$^*[\mtot=-1]$ with a global phase of $-1$, therefore, it is not invariant under this reflection, implying that, $\mu_y\ne 0$ during the propagation.

\section{Conclusions}
\label{sec:conclusions}

In this work we performed a study of the interaction of linearly and circularly polarized attosecond pulses with the first excited state of atomic neon,   Ne$^*(1s^22s^22p^5[\doubletpodd_{1/2}]3s[^1P^o])$. We made use of the alignment of this state to show the photoelectron spectra corresponding to each initial total magnetic quantum number and final state, and concluded that the electron-electron correlation plays a major role in the photoionization process. Indeed, we found that ionization channels that are only accessible due to electron correlation may dominate over channels that are open without electron correlation. In particular, our results suggest that a less unequal energy sharing between the energy that goes into excitation of the final ion and the energy carried by the outgoing electron is favored over a more unequal energy sharing, even when correlation is required for the former to occur. Besides, we also analyzed the PMDs induced by circular polarized pulses in the single photon absorption regime, illustrating that the electrons suffer a drift in the direction of the rotation of the polarization of the laser which is caused by the interference of the accessible ionization channels.

Finally, we discussed the dynamics of the electronic cloud during the laser pulse, showing that some states exhibit an asymmetry in the component of the dipole moment parallel to the propagation of the laser. This means that a circularly polarized laser pulse can be used to distinguish the chirality of atomic states.

To carry out this study, we benefited from the versatility of the TD-RASSCF-S method, which is able to accurately describe strongly correlated many-body systems with the use of small number of orbitals. Furthermore, the efficiency of the method allows us to compute the transfer of angular and magnetic momentum among orbitals with a full-description of the many-electron wave function, that, to the best of our knowledge, has not been done before for atoms containing more than two electrons. 

This investigation constitutes a first step. In the future we will explore, for instance, the interaction of aligned excited states with bi-circular polarized or elliptically polarized laser pulses, aiming to compute time-delays in photoionization and propose experiments to measure it with high accuracy.

\begin{acknowledgments}
J.J.O. gratefully acknowledges the funding under Juan de la Cierva-Incorporaci\'on programme granted by Ministerio de Ciencia e Innovaci\'on (Spain). The numerical results presented in this work were obtained at the Centre for Scientific Computing, Aarhus.
\end{acknowledgments}

\bibliography{time_dependent_many_e}

\end{document}